\documentclass[aps,pre,twocolumn,showpacs]{revtex4}
\usepackage{graphicx}

\begin{document}

\title{Multifractal analysis of  normal RR heart-interbeat signals in power spectra ranges.}
\date{\today}
\author{Danuta Makowiec}
\email[e-mail address:  ]{fizdm@univ.gda.pl}
\affiliation{Institute of Theoretical Physics and Astrophysics, University of Gda\'nsk ul. Wita Stwosza 57, 80-952 Gda\'nsk, Poland}
\author{Rafa\l \ Ga\l \c aska}
\affiliation{1st Department of Cardiology, Medical University of Gda\'nsk ul. D\c ebinki 7, 80-952 Gda\'nsk, Poland}
\author{Andrzej Rynkiewicz}
\affiliation{1st Department of Cardiology, Medical University of Gda\'nsk ul. D\c ebinki 7, 80-952 Gda\'nsk, Poland}
\author{Aleksandra Dudkowska}
\affiliation{Institute of Theoretical Physics and Astrophysics, University of Gda\'nsk ul. Wita Stwosza 57, 80-952 Gda\'nsk, Poland}

\begin{abstract}
Power spectral density is an accepted measure of heart rate variability.
Two estimators of multifractal properties: 	Wavelet Transform Modulus Maxima
and Multifractal Detrended Fluctuation Analysis are used to investigate multifractal properties for the three strongly physiologically grounded components of power spectra: low frequency (LF), very low frequency (VLF) and ultra low frequency (ULV). Circadian rhythm changes   are  examined by discrimination of daily activity from nocturnal rest. 
Investigations consider  normal sinus rhythms  of healthy 39 subjects  which are grouped in two sets: 5-hour wake series and  5-hour sleep series. Qualitative arguments are provided to conjecture the presence of  stochastic persistence in LF range, loss of  heart rate variability during night in VLF range and its increase in ULF.
\end{abstract}
\pacs{87.19.Hh, 87.80.Vt, 89.75.Fb, 05.45.Pq}
\maketitle

\section{Introduction}
Power spectral analysis of  normal RR- series of heart rate has gained popularity  as a simple, non-invasive measurement capable of assessing dynamic changes in the autonomic nervous system's control of heart rate \cite{Guide}. Power spectrum means frequency analysis to identify superimposed oscillations which contribute to variations in heart rate. It has been observed \cite{fractal} that in the very low frequency range, from 0.0003 to 0.04 Hz, the power spectral density decays as $1\slash f^\beta$. This property provides  evidence of long range dependences in the data which, when viewed within the right range of scales, leads to the notion of fractality or self-similarity: statistical properties  measured at some time scale look just like an appropriately scaled same statistical properties measured over a different time scale. 

However long range dependences leave the small-time scale behavior  (high frequency and low frequency components of the spectrum) essentially unspecified. In order to be capable of accounting for small time scales we need methods which deal with regularity, it is with   differentiability, of stochastic processes. Traditionally to quantify and qualify local irregularities  the spectrum of H\"older exponents, called multifractal spectrum is indicated \cite{Riedi,FracMath}. The practical estimations of exponents is performed in three steps. First, from the observed time series $X(t)$ one computes  multiresolution coefficients $T_X(a,t)$, quantities that depent  jointly on time $t$ and  an analysis scale $a$. Second, one constructs the partition function $P(q,a)$ from these coefficients weighed by q powers. Third, one measures the slope $\tau(q)$ in $\log a$ of versus $\log P(q,a)$ diagram. The multifractality of $X(t)$ means that $\tau(q) \neq qH$.

The aims of the following  are:

--- To provide links between properties learned from multifractal studies and characterization of the heart control system  obtained by power spectral density analysis  in physiologically grounded frequency intervals. Especially, daily activity will be compared to nocturnal rest. By separating wake from sleep human stages we want to minimize effects related to nonstationarity in a signal. So far  the problem whether the multifractal properties depend on a man activity is discussed and  contradicting conjectures are suggested  \cite{DayNight,Kiyono2005}.

--- To present in a compact and uniform way the numerical techniques used in discovering multifactal properties. We discuss results arisen from  two most popular methods:  Wavelet Transform Modulus Maxima \cite{BacryMuzyArnedo} and Multifractal Detrended Fluctuation Analysis \cite{MDFA}. Moreover, following \cite{Lashermes},  we consider the linearization effect in scaling of a structure function $\tau(q)$ by  searching for points where  a multifractal spectrum concentrates.

The estimates are applied to two types of series: to  a normal RR-signal extracted from ECG recordings and to the series obtained by integration of a normal RR-signal. 
The multifractal properties are statistical features. They are obtained  numerically following heuristic procedures. So that results could depend on details of both - properties of series and numerical procedures. Therefore, for reliability of the discussed multifractality it should be important to confirm  results in various ways \cite{Oswiecimka}.

The paper is organized as follows. In Sec.II we give a brief physiological background to reasons of heart rate variability. In Sec.III the multifractal formalism is introduced and discussed. Then, Sec. IV, we shortly describe data which is used in our investigations. Sections V and VI contain the results: the group partition functions together with  variety of linear fits which can be found to these functions (Sec. V) and the multifractal spectra estimated in  physiologically grounded time intervals. The last section, Sec. VII, presents suggestions for further work. 
We believe that our considerations, though concentrated mostly on the purely descriptive aspect of multiscaling,  are important in putting forward the research aimed on reasons for complex dynamics that is observed by the heart rate.

\section{Heart rate variability by power spectrum analysis}
\begin{figure}
\includegraphics[width=0.25\textwidth]{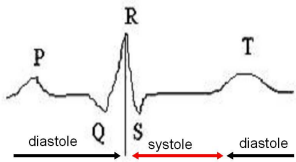}  
\caption{\small ECG diagram for a single heart period. Subsequent elements corresponds to atrial depolarization: P-wave, ventricular depolarization: QRS-complex and  ventricular repolarization: T-wave.} 
\label{pqrst}
\end{figure} 

\begin{figure}
\includegraphics[width=0.45\textwidth]{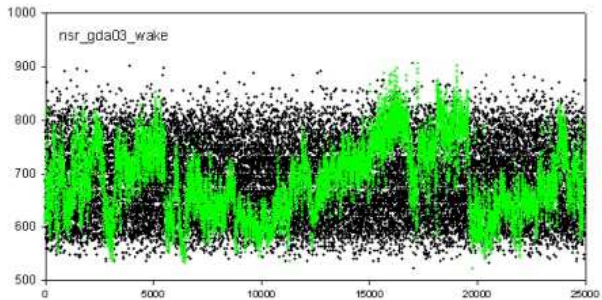}  
\caption{\small A series of RR normal heart beat intervals (green) and the same data shuffled at random. The series is 5-hour record of daily activity of a healthy subject(nsr\_gda03\_wake \cite{amgdata})} 
\label{series}
\end{figure}

The timing of a normal heart beat consists  of three elements easily recognizable in an electrocardiogram  (ECG), see Fig.~\ref{pqrst}. In a normal heart beat the following sequence of events must occur: P-wave (describing depolarization of atria), QRS-complex (associated to contraction of ventricles) and then T-wave (showing relaxation of ventricles after the contraction).
The so-called normal sinus rhythm means that each P-wave is followed by QRS-complex and the rate is characterized by 60 to 100 beats per minute with 10\% variation \cite{Guide}.
The time intervals between successive R-peaks of QRS-complex are called RR-series.
Heart rate variability refers to increases and decreases over time in the RR interval.
In Fig.~\ref{series}  a RR-series is compared to the same data but shuffled at random. An-eye inspection hints that there must be a relation between  consecutive heart beats.

A heart beat is generated by the autonomous pacemaker, called sino-atrial node, located at the top of the right atrium. However, the frequency is mediated by the neural (autonomic nervous system --- ANS) and hormonal (rennin - angiotensin - aldosterone system) activity \cite{Guide,Jackson}. This activity is controlled by large number of different feedbacks acting simultaneously. For example, both  the blood pressure as well as  changes in the blood pressure  stimulates  baroreceptors  \cite{Malik},  and baroreceptors generate a signal to ANS. 

ANS consists of two parts which act antagonistically --- the sympathetic subsystem  activity accelerates the heart rate and enhances the heart muscle contractility while the parasympathetic subsystem activity elongates the interbeat intervals. This permanent conflict is supposed to be the reason for  complex fluctuations which the normal heart rate displays. It is a widely accepted conjecture \cite{Guide,Malik,Gregory,Jackson,Sandercock,Frenneaux,Jong}, strongly supported by physicists \cite{Ivanov,filter, Kotani, Kiyono2005, Makowiec,Physica04} that variability of the heart rate reveals the delicate balance between the two branches of ANS and even more ---  $1\slash f$ scaling in healthy heart rate requires the existence of  this balance \cite{Struzik}. 

A recent introductory to the methodology of the heart rate variability can be found in \cite{Jong} and discussion of autonomic system influence in \cite{Frenneaux}. Here we concentrate on the  so-called frequency-domain measures resulted from analysis of power spectral density of a RR tachogram \cite{Guide}. To avoid nonstationarities short-time recordings (several minutes) are used. Three main spectral components are distinguished in this spectrum: very low frequency: $0.003<f<0.04$ Hz (VLF), low frequency $0.04<f<0.15$ Hz (LF) and high frequency: $0.15<f<0.4$ Hz (HF) components. Longer recording (hours) spectral analysis provides an additional range $f<0.003$ Hz termed ultra low frequency component (ULF).

The physiological explanation of ULF and VLF is less defined than other components. 
The power spectral density  for these components is of power-low form $1\slash f^ \beta$. The  value of $\beta$ can discriminate healthy heart rate from a rate with heart failure \cite{fractal}. Moreover, reduced levels of the spectral power have been identified as predictions of all-cause mortality (see  \cite{Sandercock} for discussion). There is some evidence that it is influenced by thermoregulatory mechanisms and humoral rennin-angiotensin system, but physical activity also notably influences to the power \cite{Roach}. 

The higher frequencies --- HF component, are modulated predominantly by parasympathetic activity. A sino-atrial node responds to this activity at the respiratory frequency.  LF component is modulated by both sympathetic and parasympathetic neural influence. The ratio of the spectral power LF to HF  is used as a measure of sympathovagal balance. However this measure is found to be controversial \cite{Jong}.

It is known that multifractal analysis can serve indicators of a state of ANS. The definite reduction of multifractal properties, namely the  transition from multifractal to monofractal, is observed in multifractal picture of ECG of patients with congestive heart failure \cite{Ivanov,filter}. The  modifications in multifractal spectra are observed in case of other  heart diseases \cite{Kotani,Makowiec,MeyerStiedl}. Following those investigations  some heart rate variability measures have been proposed, e.g , spectrum width, spectrum location, curvature of a spectrum. Especially, it has been  observed that a multifractal spectrum is distinct from a monofractal plot by existence of points (more than one) at which the spectrum is concentrated  \cite{Makowiec}.  Speculations are continued that multifractal analysis offers potentially promising tools for heart rate variability assessments, but standards are lacking.

\section{Multifractal approach}
Multifractal analysis of time series is strongly mathematically grounded, see for example \cite{FracMath,Riedi}. Roughly speaking by multifractal study one wants to express relations between points $X(t')$ and  a given point $X(t)$, where $t'\in {\cal N}(t) $ belongs to some neighborhood of $t$,  by comparing a part of series to a polynomial with a singularity exponent $h(t)$:
\begin{equation}
|X(t') -X(t)| \sim K |t'-t|^{h(t)} \qquad {\rm for}\quad t'\in {\cal N}(t)
\end{equation}
The singularity exponent $h$ is often called H\"older exponent.  Typically a process $X(t)$ will posses many different singularities when moving $t$. The multifractal spectrum measures the frequency of occurrence of a given singularity exponent. Since  singularity values change along the series vividly then the probability of occurrence is measured by the Hausdorff dimension of the subset of time where the same singularity exponent value is accounted:
\begin{equation}
D(h)= \dim_H \{t: h(t) =h \}
\end{equation}
However transferring these ideas to numerical estimates of a series which arises from some natural phenomena is not obvious. Both tasks: extracting singularity exponents and evaluating the Hausdorff dimension, are numerically difficult and calculations cannot be performed automatically.  

\begin{figure}
\includegraphics[width=0.45\textwidth ]{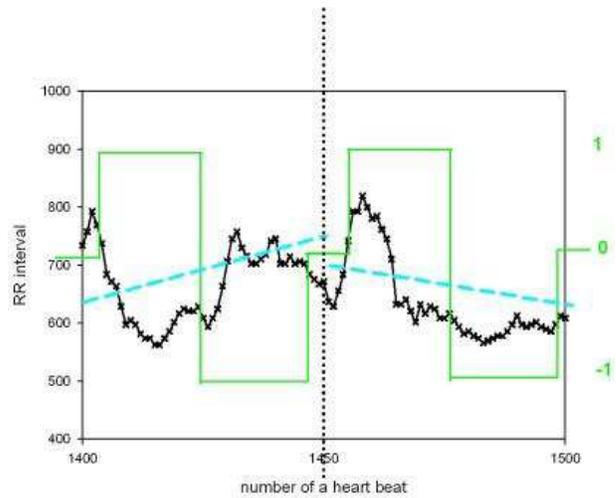}  
\caption{\small WTMM vs MDFA methods in a simplest form. A signal is divided into boxes of size 50 heart beats. In each box, the oscillations of a signal are represented by two estimates: a wavelet coefficient (here, the Haar wavelet is plotted by a solid line) and a fluctuation from a main trend (a dashed line plots the linear trend for a given box). } 
\label{methods}
\end{figure}

When one deals with discrete time series then special difficulty relates with determination of the neighborhood  $ {\cal N}(t)$ . In case of continuous data $x(t)$, the fractal singularity is estimated with respects to the small enough $|t-t'|$ interval. But how to transfer this condition to a discrete signal? At the smallest possible interval, namely a unit interval, each two points can be connected by any function. If a regular function is chosen then the fractal dimension will be zero. Does it mean that the described procedure is useless?  In Fig.\ref{methods} we present a piece of  series consisting of a hundred points which are divided into two equal size boxes. The question is whether the neighborhood consisting of 50 points or more can be considered as small enough to provide estimates for roughness of a series. In our opinion any scale is appropriate to discuss the fractal properties of a discrete time series. Following this belief we will continue to call our search as calculations of multifractal properties.

Both estimates: singularity exponent $h$ and its fractal dimension $D(h)$ can be found thanks to  scaling properties of the partition function  and their relation to  a global characterization of a fractal object  \cite{FracMath,ArnedoBacryMuzy}. 

Basically, there are  two approaches to construct the multiresolution coefficients: \\
(1) aggregation of a signal within a $K$-th box of a given size $n$, i.e., $ X(n,K)= {1\over n}\sum_{i=1}^n X(Kn+i)$, \\
and \\
(2) increment of  a signal within a $K$-th box of a given size $n$, i.e.,  $X(n,k)=X(Kn+n)-X(Kn)$. 

By the wavelet technique and detrended fluctuation method one deals with both:  increments between carefully chosen averages are estimated. Having a signal divided into boxes of a given scale, see Fig. \ref{methods},  the wavelet transformation in a $K$-th box: $T_\psi(n,K)$  measures the signal local oscillations inside a box. One can also estimate signal oscillations by measuring  departures $F_{P_m^K}(n,K)$ of the signal from a  polynomial  trend $P_m^K$ of order $m$ found for points of $K$-th box. Finally, the partition function for a given scale is defined as the sum of either  selected --- local absolute value maxima in case of wavelets, or all --- in case of detrended fluctuations, oscillations weighted by $q$ exponent. Both methods are heuristic and therefore below we give the precise  mathematical formulas of estimates.

Let us assume that a series $\{X(i)\} $ is divided into boxes consisting of $n$ points. Then the following calculations lead to the partition functions $Z(n,q)$ and $F(n,q)$, respectively.

--- In Wavelet Transform Modulus Maxima (WTMM) method  a multiresolution coefficient is obtained as follows, \cite{BacryMuzyArnedo}: 
\begin{equation}
T_\psi(n,K)=\sum_{i\in K} X(i) \psi_K(n,i)
\end{equation}
where $\psi_K(n,i)$ is the mother wavelet transformed to scale  $n$ and shifted to $K$th box. Then  the partition  function $Z(n,q)$ is:
\begin{equation}
Z(n,q) = \sum_{K: \sup { local\ maxima \atop across\  finer\ scales}} |T_\psi(n,K)|^q  
\label{Z}
\end{equation}
where sum takes supremum over so-called maxima lines, i.e., lines which link the local maxima at the present scale with the maxima at the previous scales. 

--- In  Multifractal Detrended Fluctuation Analysis(MDFA) approach a multiresolution coefficient is defined as, \cite{MDFA}:
\begin{equation}
S_{P_m^K}(n,K)=\left(\sum_{i\in K} [X(i) -{P_m^K}(i)]^2\right)^{\frac{1}{2}}
\end{equation}
and then the partition function $S(n,q)$ is given as
\begin{equation}  
F(n,q) = \sum_{K} S_{P_m^K}(n,K)^q  
\end{equation}

Let us notice that the supremum over the maxima lines, calculated in WTMM approach (\ref{Z}), provides the inter scale links while MDFA method works on each scale independently.

Dependence on $n$ of any partition function $P$ leads to  so-called structure function $\tau (q)$ as follows:
\begin{equation}
P(n,q)\sim n^{\tau (q)}
\end{equation}
and then by applying  a Legendre transform to a curve $ (q,\tau(q)) $ one obtains the multifractal spectrum $(h,D(h))$: 
\begin{equation}
 h=\frac{d\tau (q)}{d q} \qquad D(h)=qh-\tau(q)
\label{Legendre}
\end{equation}
which links the singularity exponent value $h$ with its probability to appear $D(h)$ in a signal.

In case of a monofractal signal one expects a linear structure function $\tau(q)$ and then a corresponding spectrum should consists of  one point $(h,D(h))$. But the numerics overestimate the real H\"older exponent $h$. Therefore the numerical results are sometimes called  the coarse H\"older exponents \cite{Riedi}. Additionally, the Legendre transform  (\ref{Legendre}) significantly magnifies any small departures from linearity. Finally, what we obtain from numerical estimates are spectra which are far from being a point-like \cite{Makowiec,Oswiecimka}. However the maximum of a spectrum curve is directly related to the theoretical spectrum point and the maximum point is the highest occupied point of the whole spectrum \cite{Makowiec}. A multifractal spectrum is distinct from a monofractal plot by existence of many highly occupied points. This feature is called the linearization effect  and is suggested to be related with the nature of the process \cite{Lashermes}.

\begin{figure}
\includegraphics[width=0.49\textwidth]{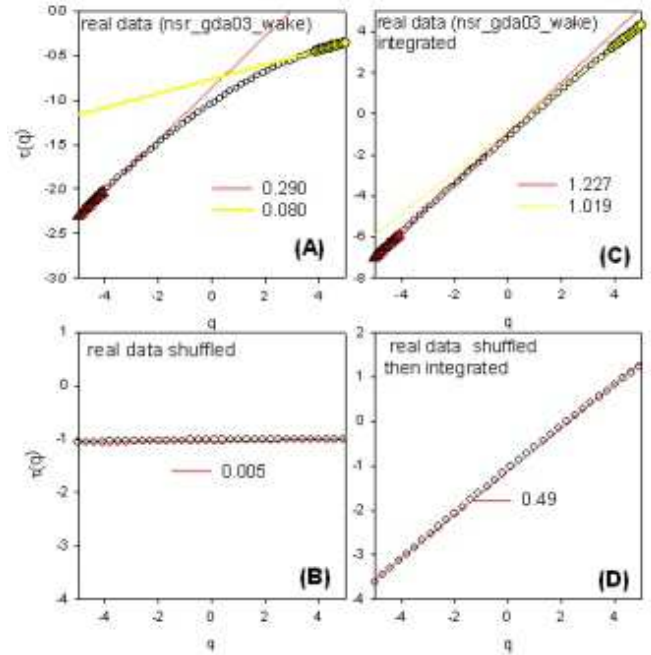}  
\caption{\small Structure functions $\tau (q)$ obtained by MDFA for a  RR signal. (A)   $\tau (q)$ for nsr\_gda03\_wake. (B) $\tau (q)$ for nsr\_gda03\_wake shuffled at random. (C) $\tau (q)$ for nsr\_gda03\_wake integrated.(D) $\tau (q)$ for nsr\_gda03\_wake shuffled at random and integrated. All linear fits provides $r^2$-errors greater than 0.98}
\label{shuffled}
\end{figure}

In general when studying empirical data what we know is that negative moments, i.e., when $q <0$, are responsible for enhancing small fluctuations while positive moments, $q>0$, amplify large fluctuations. The right wing of a spectrum corresponds to negative moments and the left wing reflects properties collected for positive moments.  

In Fig.\ref{shuffled}A we present the structure function found due to MDFA method to a series shown in Fig.\ref{series}. The structure function $\tau(q)$ of the pure signal is concave and any linear fit is loaded by a large $r^2$-error. However, one can easily see that for $|q|>4$ line fits are justified (the linearization effect) what will result in emergence of two dense occupied points on $(h,D(h))$ plane: at $h=0.290$ and at $h= 0.005$. The signal is multifractal. If the data is  shuffled at random,  then the corresponding partition function becomes a flat line, see Fig.~\ref{shuffled}B. The same result is obtained for a white noise --- a canonical example of a monofractal signal. Its multifractal spectrum consists of a point $(0,1)$. 

In the original detrended fluctuation analysis one deals with so-called profiles, see \cite{DFA}. The profile can be thought as a signal obtained after integrating series. Dealing with series of integrated RR intervals  one introduces a kind of smoothing since only growth in a data set is possible. Moreover, integrated series allows to study strength of independency in a series. In Fig. \ref{shuffled}C  the structure function obtained for the integrated series of nsr\_gda03\_wake is shown. Again the linear fits are provided for $|q|>4$. The linear approximations in these intervals give: 1.227 and 1.019 what locates the multifractal spectrum between these two points. Hence, the spectrum is moved right by about of 1 from the spectrum obtained from the pure series on the $(h,D(h))$ plane. However, if the data is independent then, see  the structure function of integrated shuffled data in Fig. \ref{shuffled}D, the linear fit provides the slope of $0.49$ what means that the spectrum is moved right but only by 1\slash 2.

\section{Heart interbeat interval data}
The series called {\bf nsr\_gda} were collected  from 39 healthy individuals (4 women, 35 men, the average age is $54 \pm 7$) without past history of cardiovascular disease, with both echocardiogram and electrocardiogram in normal range. The left ventricle ejection fraction was normal (mean LVEF $= 68.0 \pm 4.7 \%$). 
For each person the 24-hour ECG Holter monitoring was performed. The signal was digitized using Delmar Avionics recorder (Digitorder) and then analyzed and  annotated using Delmar Accuplus 363 system (fully interactive method) by an experienced physician. The minimum number of qualified normal sinus beats required for the signal to enter the study is $85\%$.
Moreover, the standard filters \cite{filter} have been applied to smooth out the data. 

From each signal the two  5-hour continuous subsets were extracted:  diurnal and nocturnal.
By extracting such subseries we wanted eliminate at least some sources of nonstationarities.
The subsets were extracted manually, with regards to two easily recognizable stages: daily activity and sleep state. In most cases these periods are consistent with hours 15:00--20:00 for the daily activity and 0:00--5:00 for the sleep state.  In the following we will denote the subseries as  nsr\_gda\_wake,  nsr\_gda\_sleep, respectively.
All RR-series and their subseries are accessible on request \cite{amgdata}.

Since the WTMM method requires data sets longer than $2^{14}$, then some of the sleep subseries have to be excluded from WTMM calculations. So that nsr\_gda\_sleep group consists of 34 data sets in case of WTMM calculations. 

\section{Scaling properties of partition functions }

Both partition functions $Z(n,q)$ and $F(n,q)$ are calculated by tools accessible from PhysioNet \cite{physionet}.  Namely, we work with  two Physionet packages: multifractal.c (prepared by Y. Ashkenazy) and dfa.c (prepared by J.Mietus, C-K Peng, and G. Moody ) adapted by us  and accessible from \cite{amgdata} 

For each individual series, regarded as  either directly as a sequence of RR intervals,  or  as a sequence of integrated RR intervals, the partition function was found for all $q$ in $[-5, 5] $ with a step $0.1$, for scales from $n= 10$ to almost $1000$ heart beats. In all calculations we applied the third derivative of a Gaussian as a mother wavelet in WTMM method, (thus up to quadratic polynomial dependencies in $t$ are eliminated by WTMM) and a quadratic polynomial as a detrending fit (MDFA). Such setting of parameters has been found by us as less dependent on the length of test series --- localization of  spectra is stable (WTMM) and width of spectra is stable (MDFA).

\begin{figure}
\includegraphics[width=0.45\textwidth]{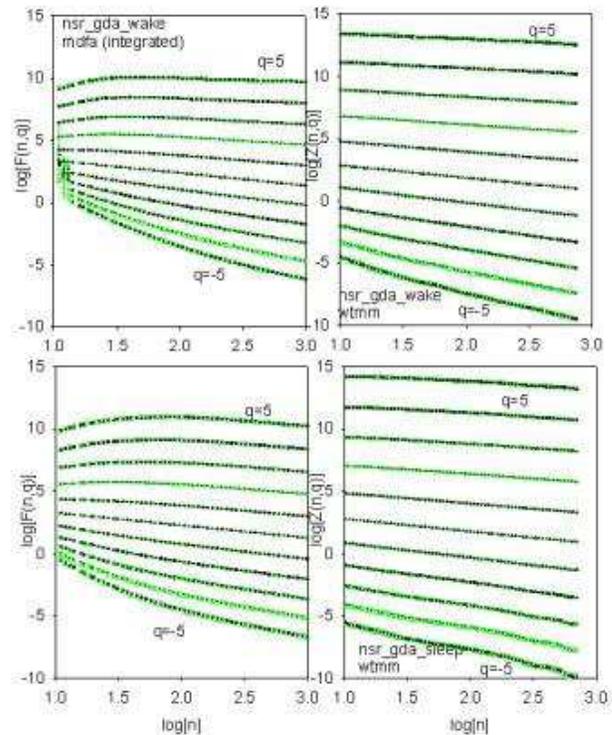}  
\caption{\small Group average of partition functions calculated by MDFA and WTMM methods for each element of the group and for different $q=-5, -4, \dots, 4,5$. In green,  statistical errors of estimates (standard deviation of a group $\backslash\sqrt N$, where $N$ number of elements in the group) are given.} 
\label{partition}
\end{figure} 

\begin{figure}
\includegraphics[width=0.44\textwidth]{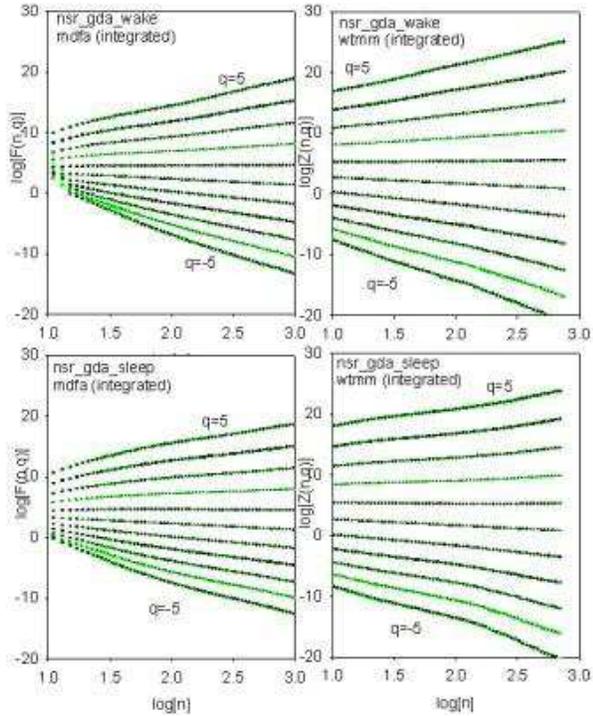}  
\caption{\small Group average of partition functions calculated by MDFA and WTMM methods for each integrated signal  of the group for different $q=-5, -4, \dots, 4,5$. In green,  statistical errors of estimates (standard deviation of a group $\backslash\sqrt N$, where $N$ number of elements in the group) are given.} 
\label{partitionInt}
\end{figure}

The partition functions $Z(n,q)$ and $F(n,q)$, shown in Figs \ref{partition}, \ref{partitionInt}, are group averages obtained from the partition functions found for each individual series by WTMM and MDFA methods, respectively. We present $Z(n,q)$ and $F(n,q)$ for  $q$-representatives $\{ -5, -4,\dots, 4,5\} $ together with their statistical errors. 
We found errors satisfactory low (with significance level $ p<0.05$ )  to discriminate values of particular $q$s. Therefore we could turn to searching for power-law scalings  of these functions. To reliably infer power-law scaling, a straight line in log-log plot has to be established. In Figs \ref{partition}, \ref{partitionInt} the log-log scale is used to enhance visually linear relations between log[scale] and log [partition function].

\begin{figure}
\includegraphics[width=0.235\textwidth]{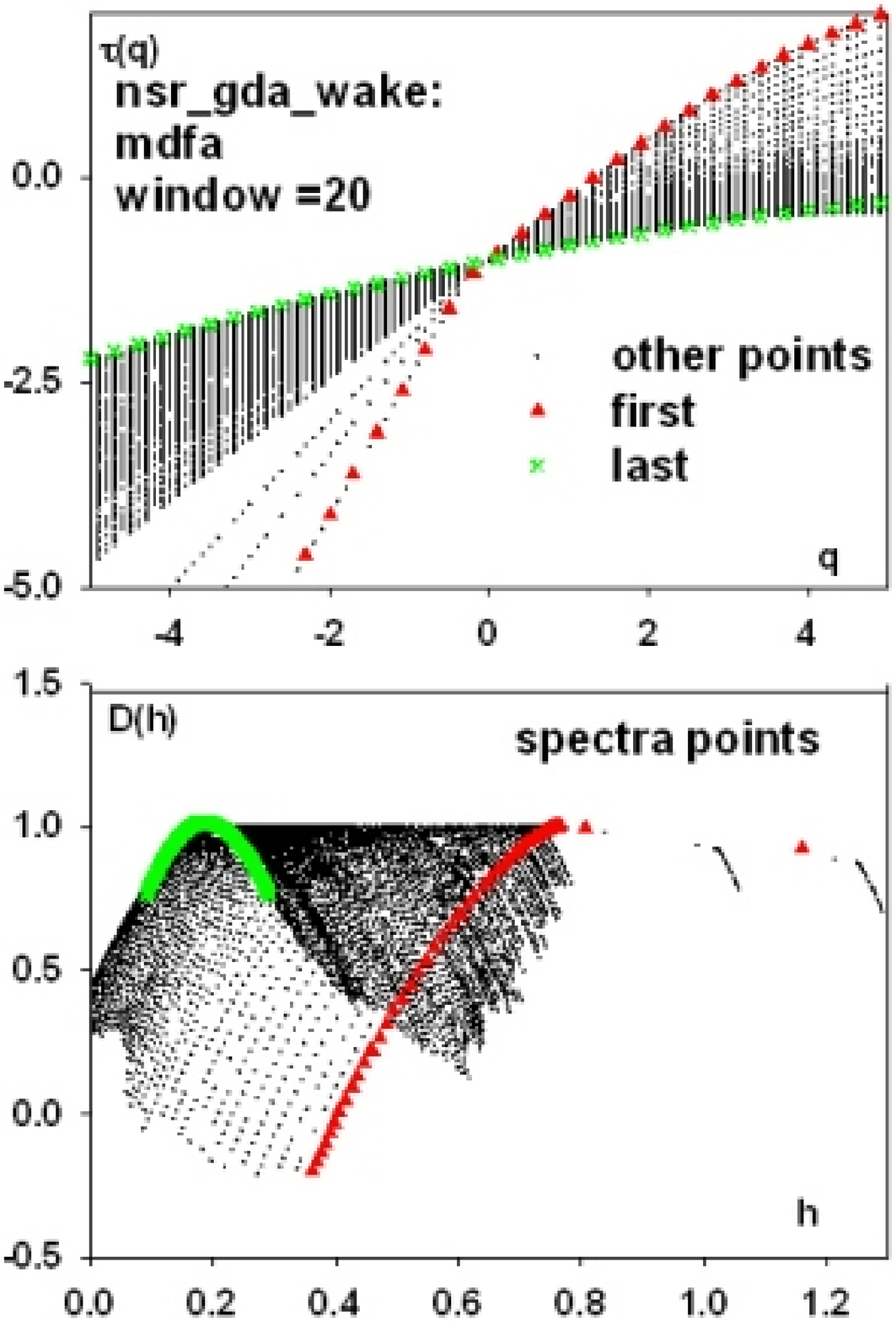}  
\includegraphics[width=0.235\textwidth]{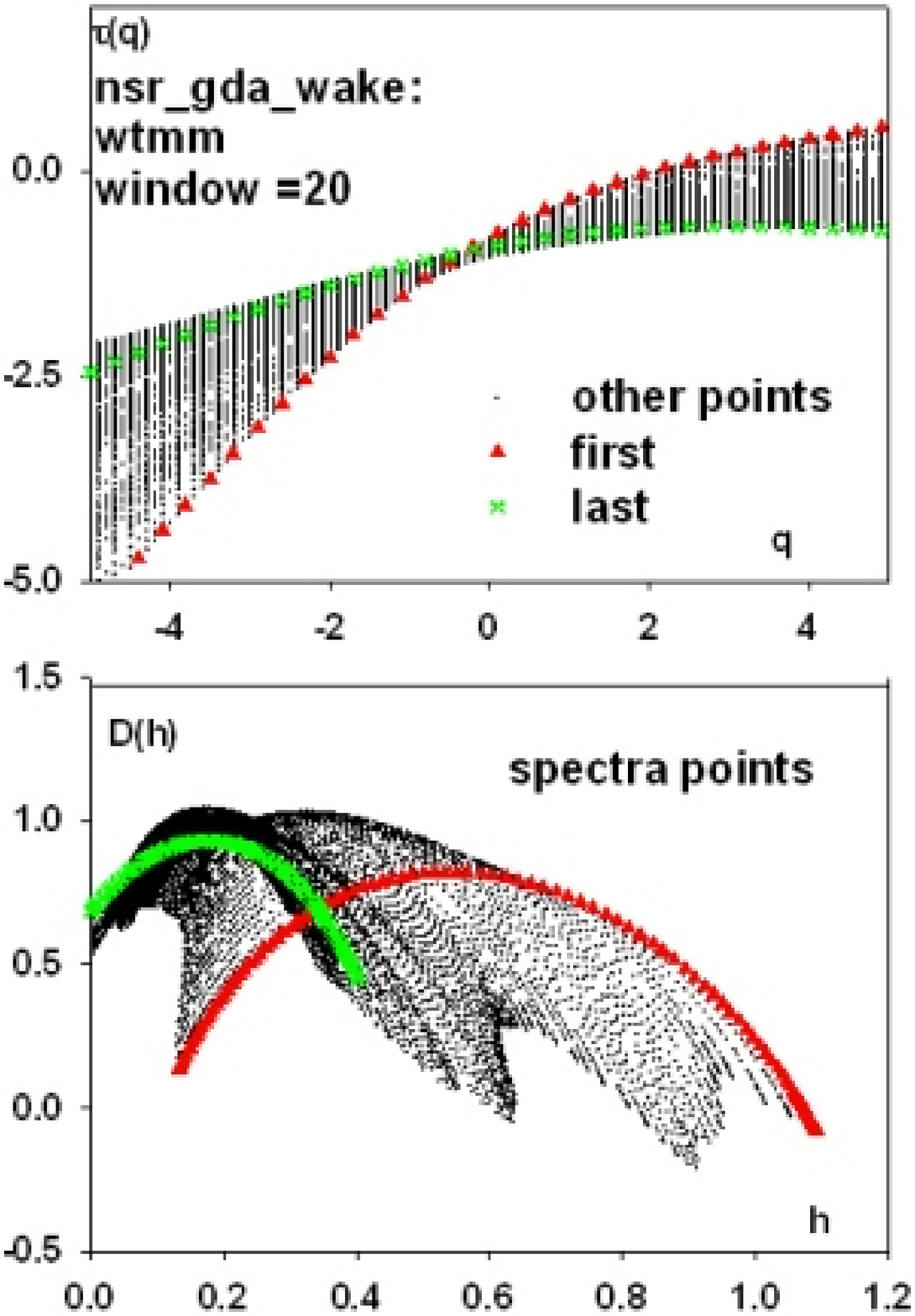}  
\includegraphics[width=0.235\textwidth]{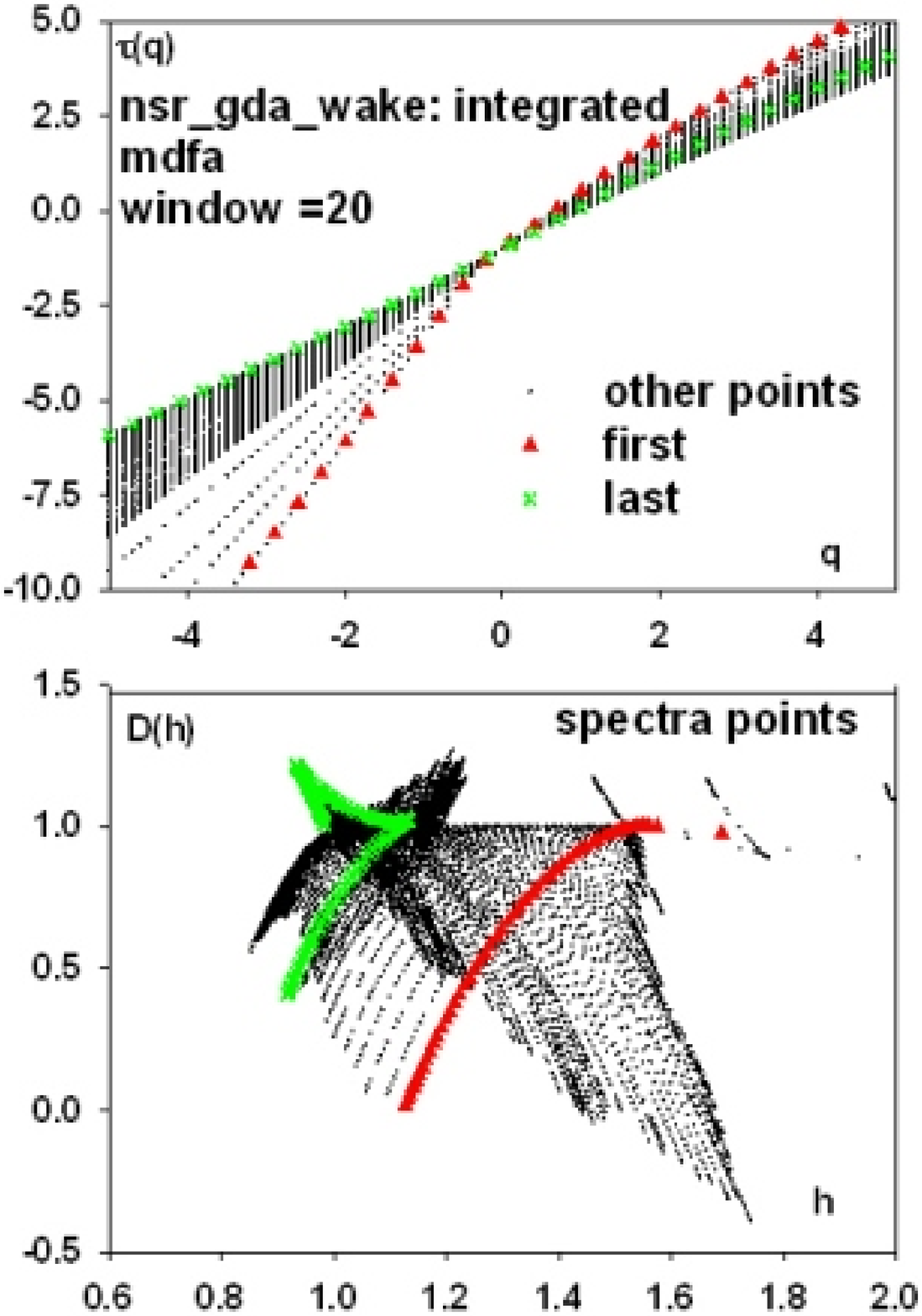}  
\includegraphics[width=0.235\textwidth]{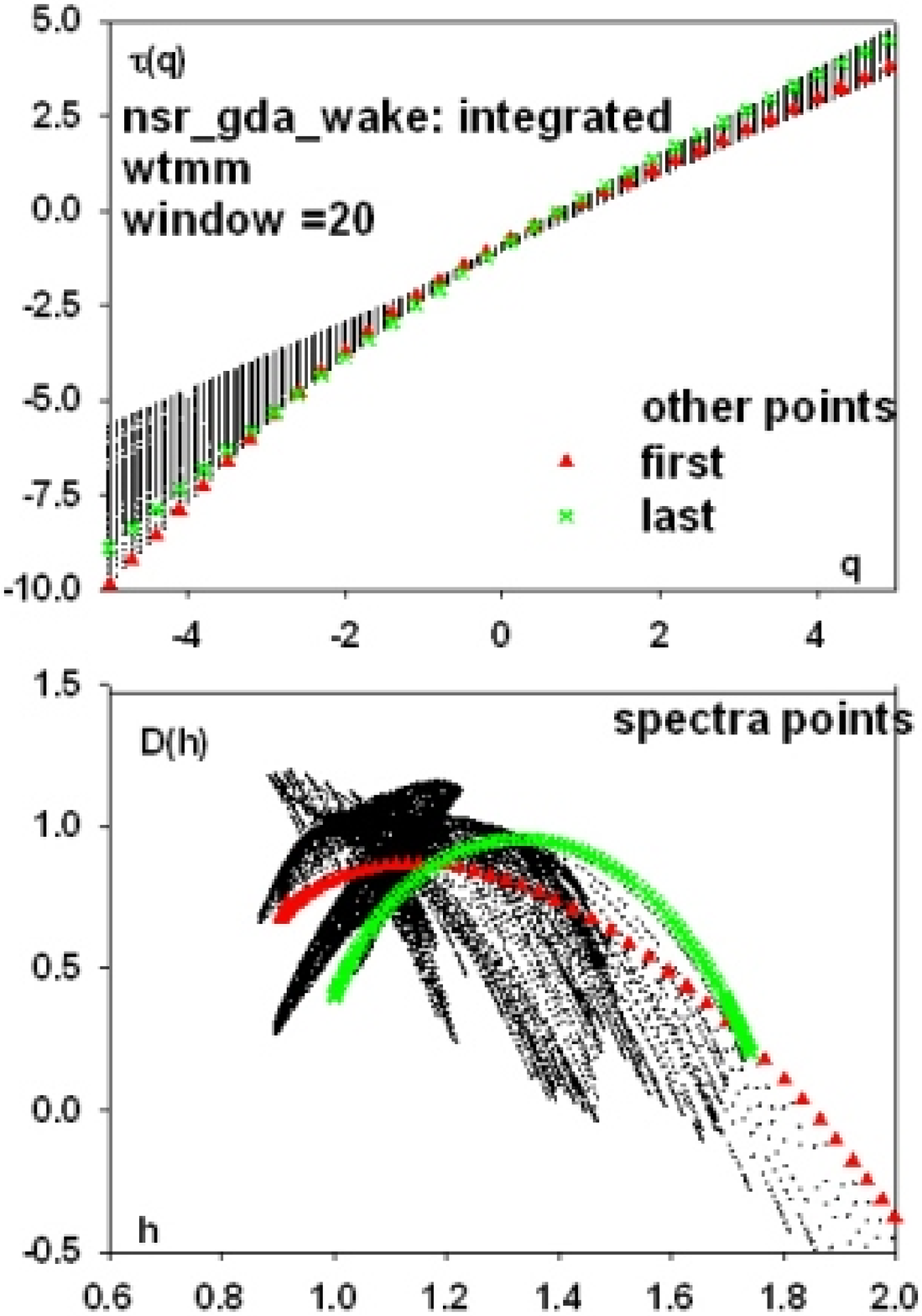}  
\caption{\small The structure functions  and spectra obtained  when the linear fits are found for subsequent 20 points of the corresponding partition functions. Group averages for integrated nsr\_gda\_wake  obtained by both methods for pure series (upper panel) and integrated series (bottom panel).}
\label{wake}
\end{figure} 

\begin{figure}
\includegraphics[width=0.235\textwidth]{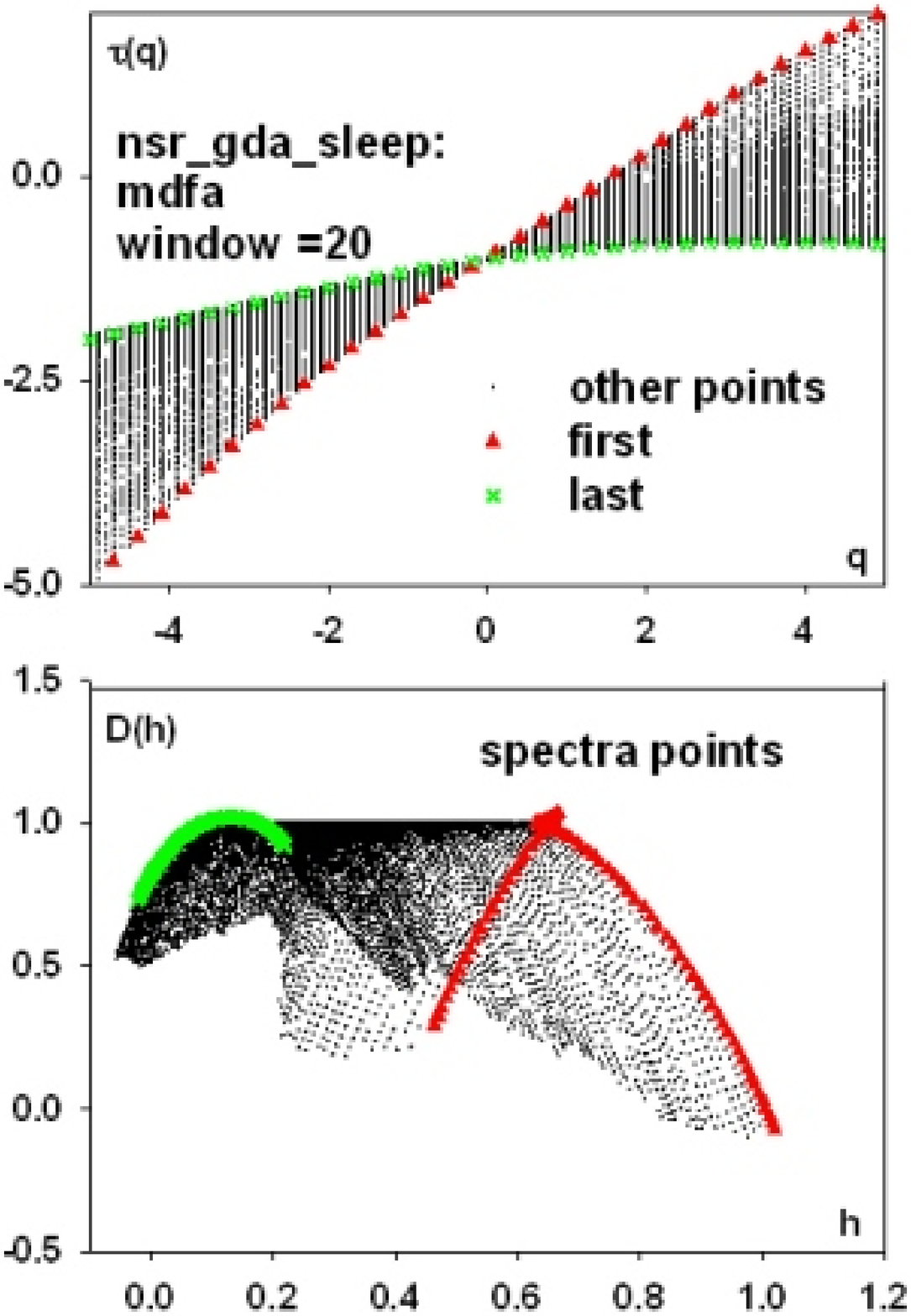}  
\includegraphics[width=0.235\textwidth]{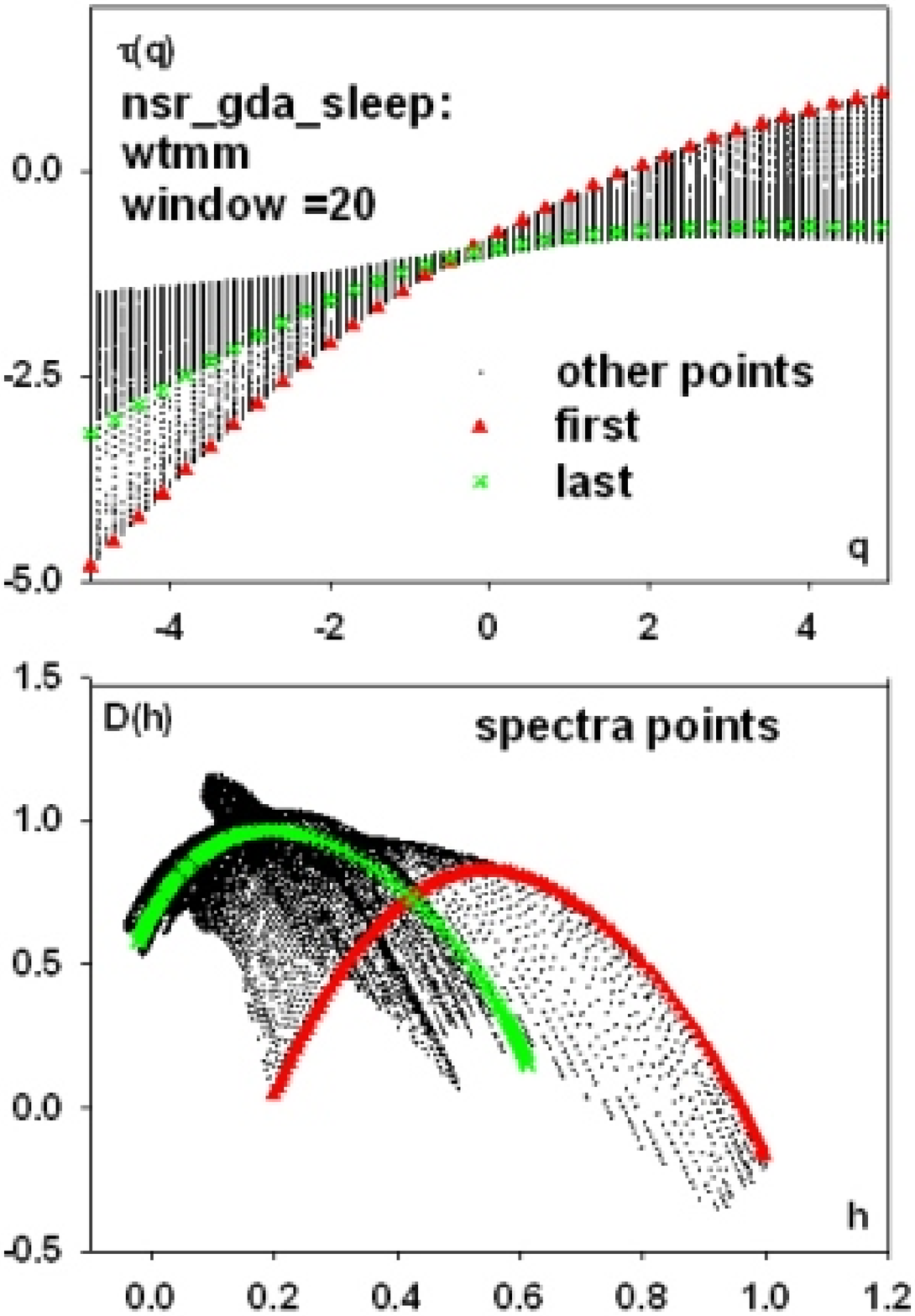}  
\includegraphics[width=0.235\textwidth]{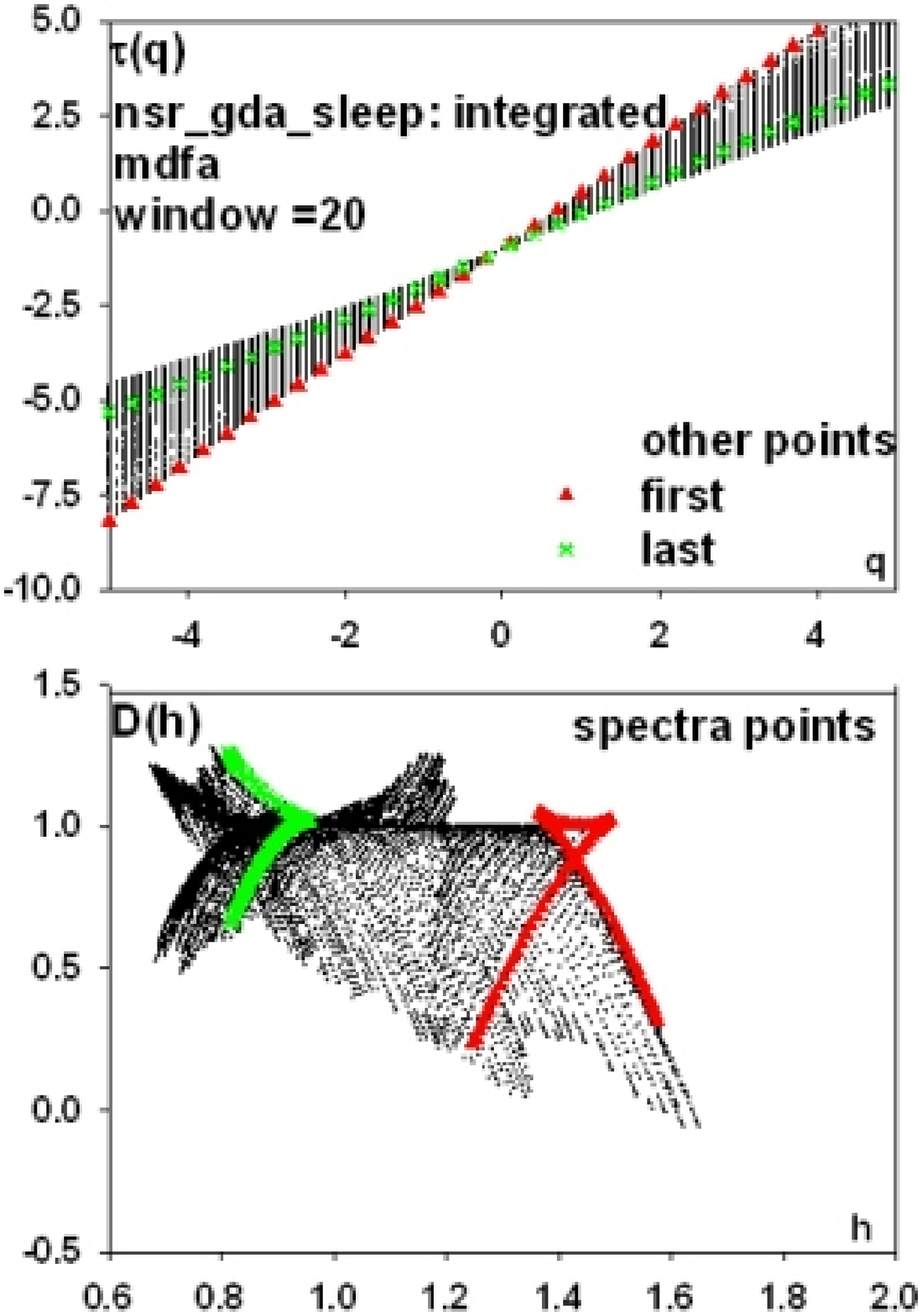}  
\includegraphics[width=0.235\textwidth]{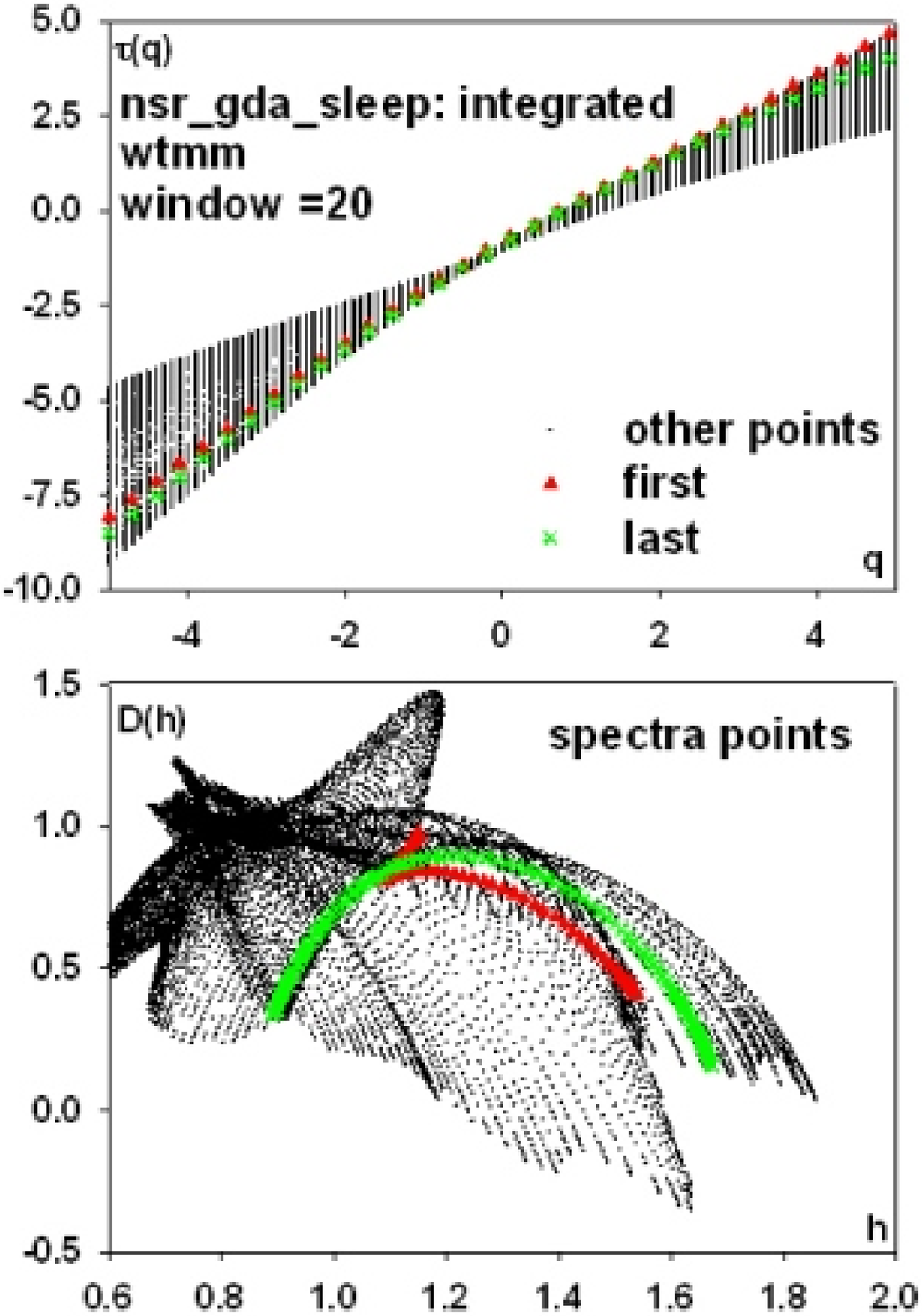}  
\caption{\small The structure functions  and spectra obtained  when the linear fits are found for subsequent 20 points of the corresponding partition functions. Group averages for  integrated nsr\_gda\_wake  obtained by both methods for pure series (upper panel) and integrated series (bottom panel).}
\label{sleep}
\end{figure}

It appears that the partition functions $F(n,q)$ change their dependence on a scale. Especially, the change is noticeable in case of pure series. Since numerical calculations are  sensitive to details of curves,  we  evaluate  subsequently local slopes of  all functions:  $\log F(n,q)$ vs $\log n$ as well as $\log Z(n,q)$ vs $\log n$ to observe changes in scalings. In particular, we calculate linear fits  for subsequent 20 points of each partition function and move with the starting point from 20th heart beat to the end of series. This way we obtain the variety of possible approximations and then we can  ask how one can estimate fine scale or\slash and large scale limits properties.

The results -- local structure functions and resulting local spectra plots are shown in Figs \ref{wake}, \ref{sleep}.  The first structure function $\tau_{first}(q)$, labeled  first (red plots) starts at the 20th heart beat. The plots labeled last in the figures  correspond to $\tau_{last}(q)$ obtained in the last possible but smaller than 1000th heart beat (green plots). One has to be aware that 20 points at fine scales correspond to about several heart beats while 20 points at large scales cover few hundred heart beats since a scale box grows geometrically with a ratio $2^{0.05}$. The spectra  are plotted below the corresponding structure functions. One can see that the 'body' of each spectrum occupies a wide part of a plane $(h,D(h))$. The strange structures appearing in some of  spectra are consequences of Legendre transformation (\ref{Legendre}) when treated by  numerical methods.  

Both methods MDFA and WTMM provide similar results in case of pure series. The fine scales are significantly separated from large scales: the first-plots are centered at $h=0.6$ while last-plots are concentrated around  $h= 0.2$. The maxima of corresponding spectra are  located similarly on $(h,D(h))$ -plane and widths of corresponding spectra are in similar relations: the fine scale widths are significantly greater than the large scale spectra. These results might suggest presence of plenty independent stochastic noises actively influencing dynamic properties of heart rate in short time scales.

The integrated series are expected to provide spectra moved right by 1 from the pure series. It appears that  this simple relation is about to be satisfied only in case of the spectra of wake series treated by MDFA method. In case of the sleep data we see that the spectrum corresponding to large scales is located  in $ h<1$ area. This observation suggests that nocturnal series are characterized by stronger independence between subsequent heart beats than diurnal series. In case of WTMM method both limits: fine and large scale provide  spectra centered  similarly at about $h=1.2$. Therefore one can conjecture that WTMM spectra of integrated series represent some average multifractal characteristics.

\section{Spectra}

\begin{figure}
\includegraphics[width=0.23\textwidth]{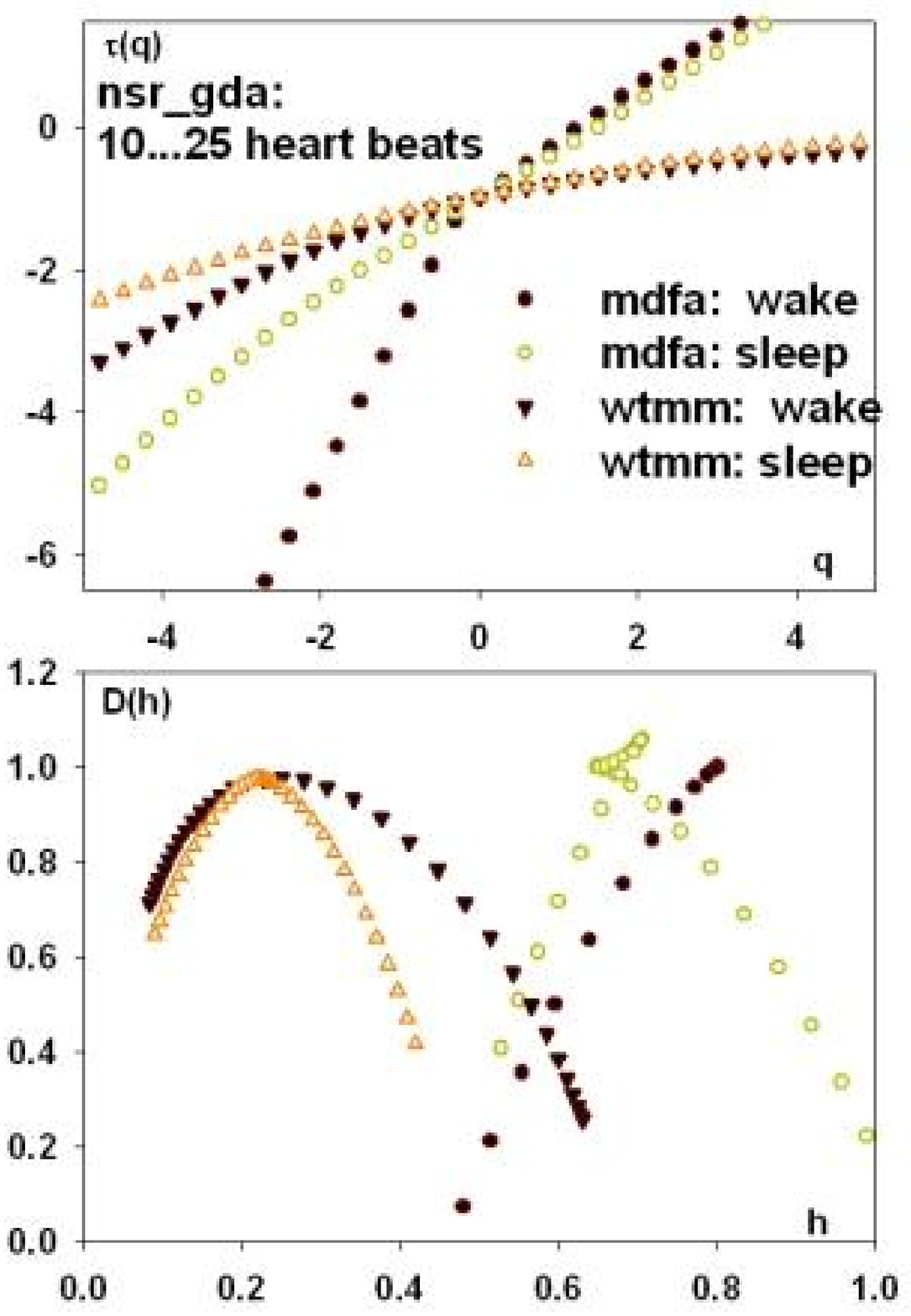}  
\includegraphics[width=0.23\textwidth]{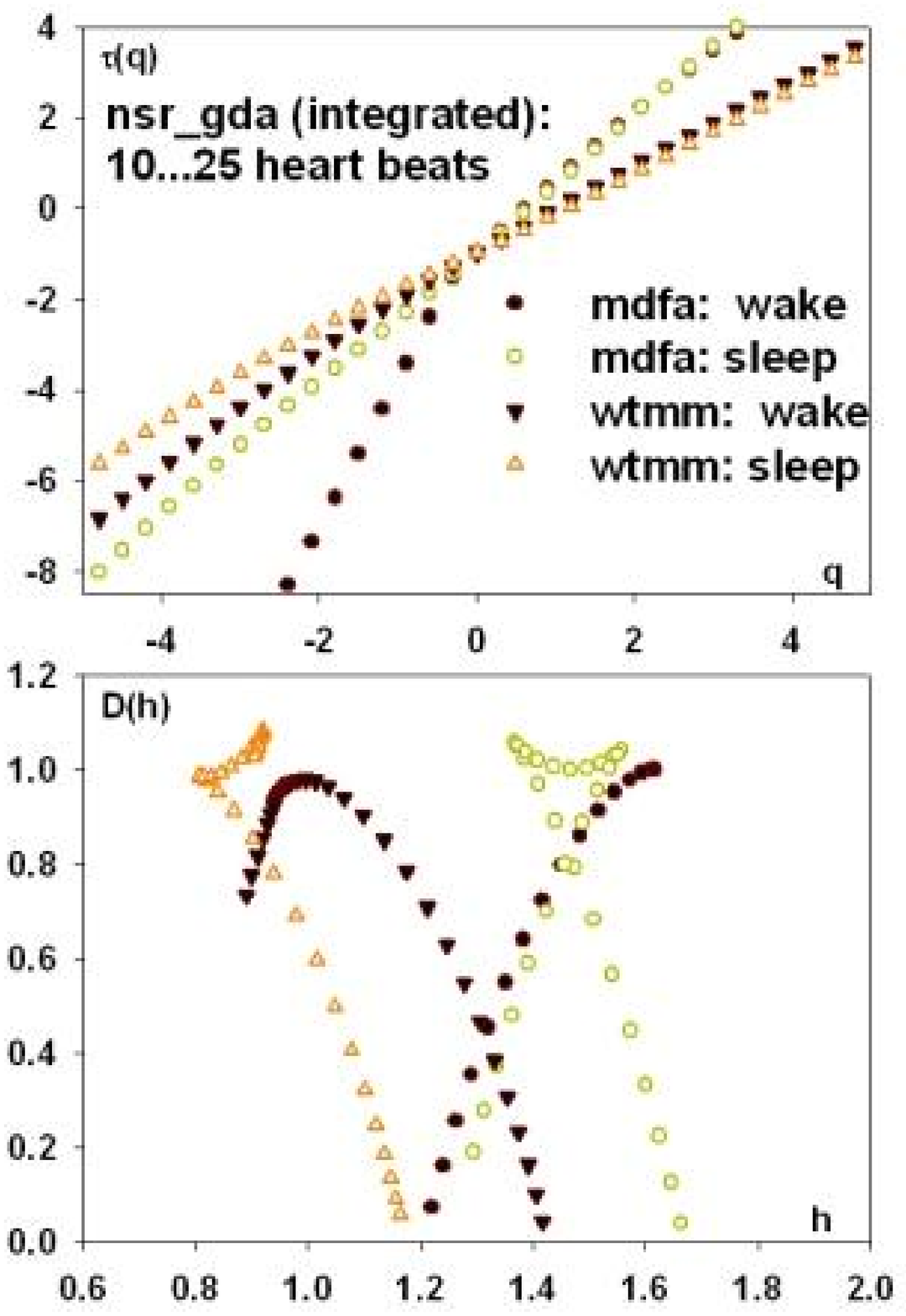}  
\includegraphics[width=0.23\textwidth]{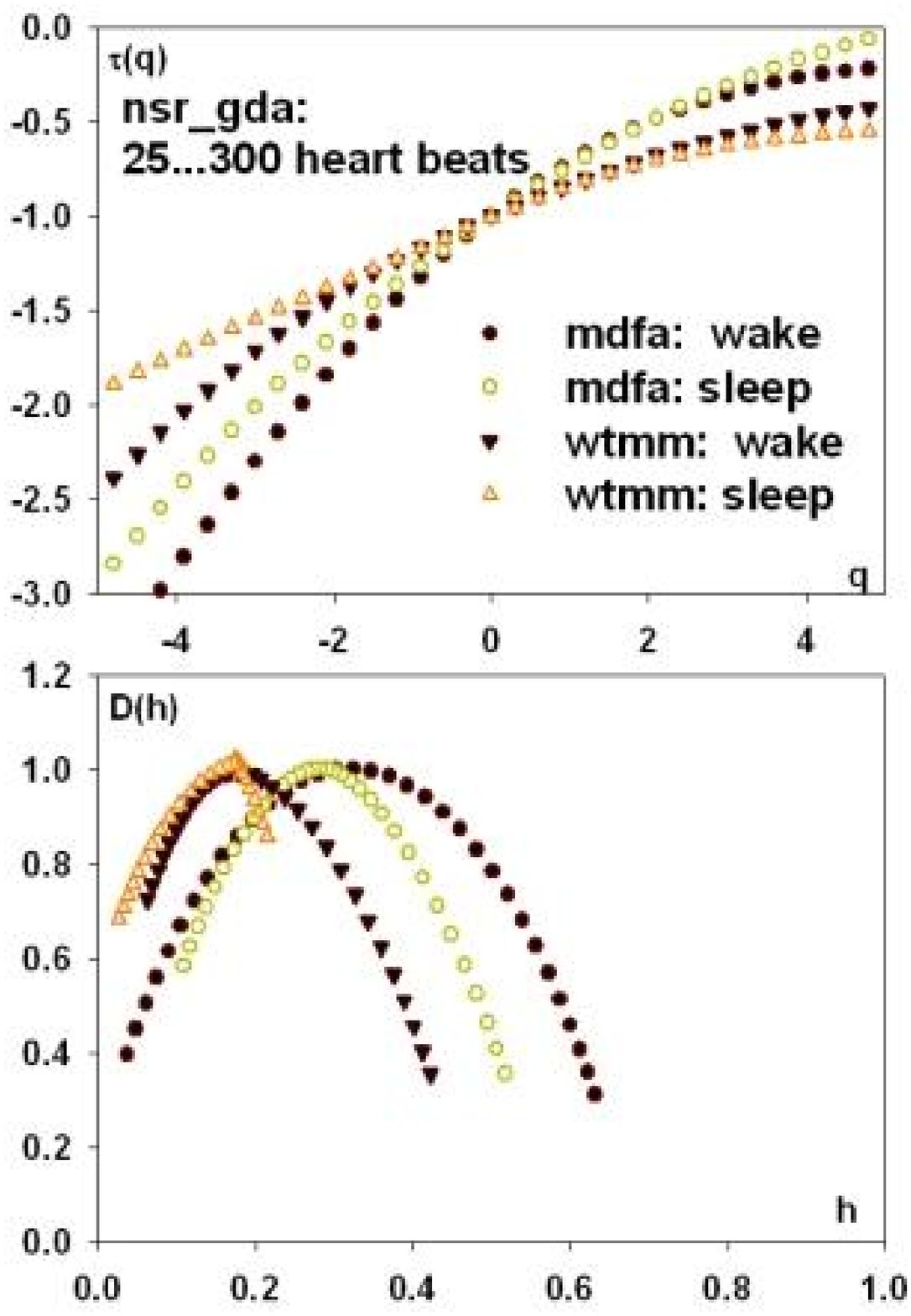}  
\includegraphics[width=0.23\textwidth]{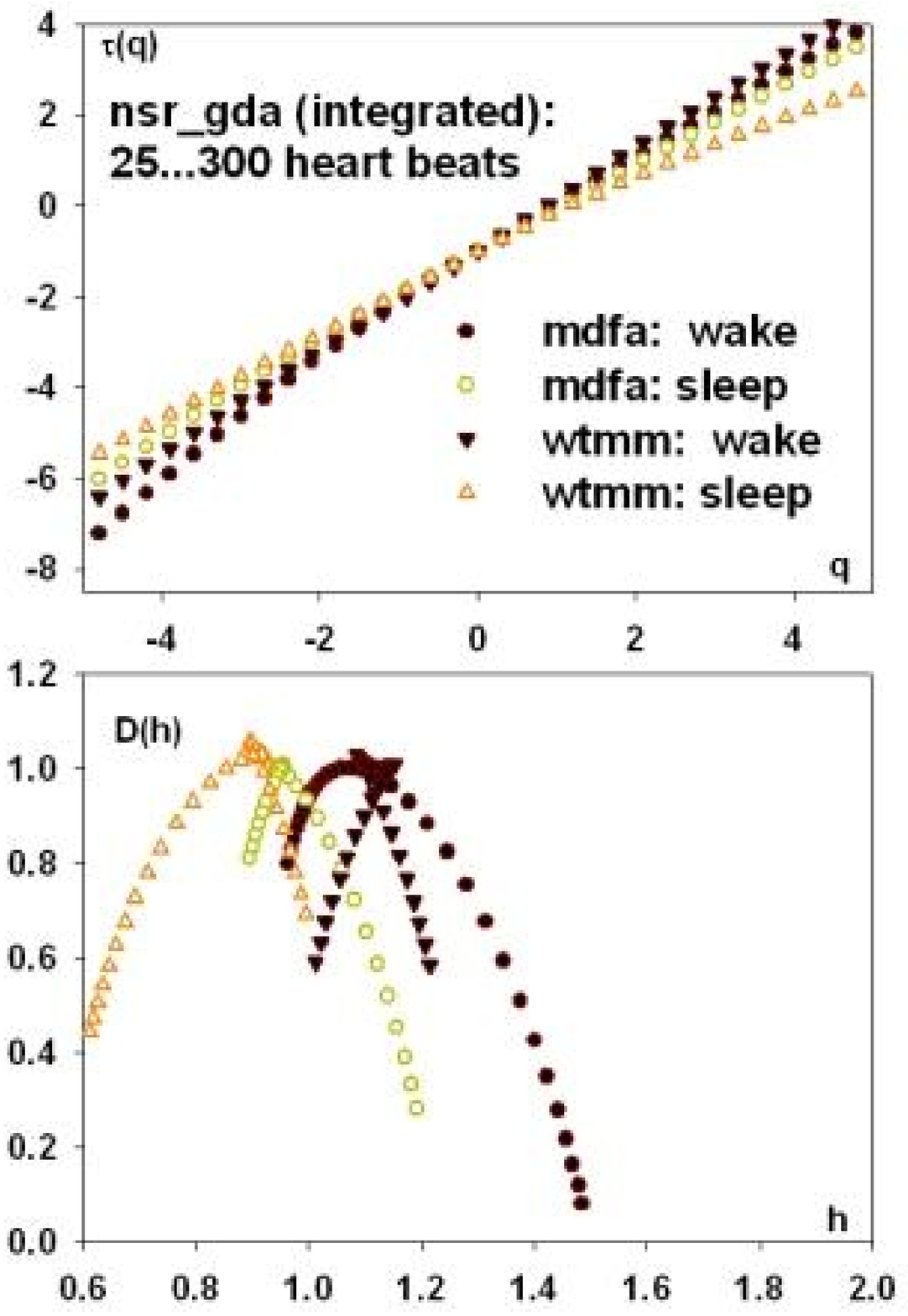}  
\includegraphics[width=0.23\textwidth]{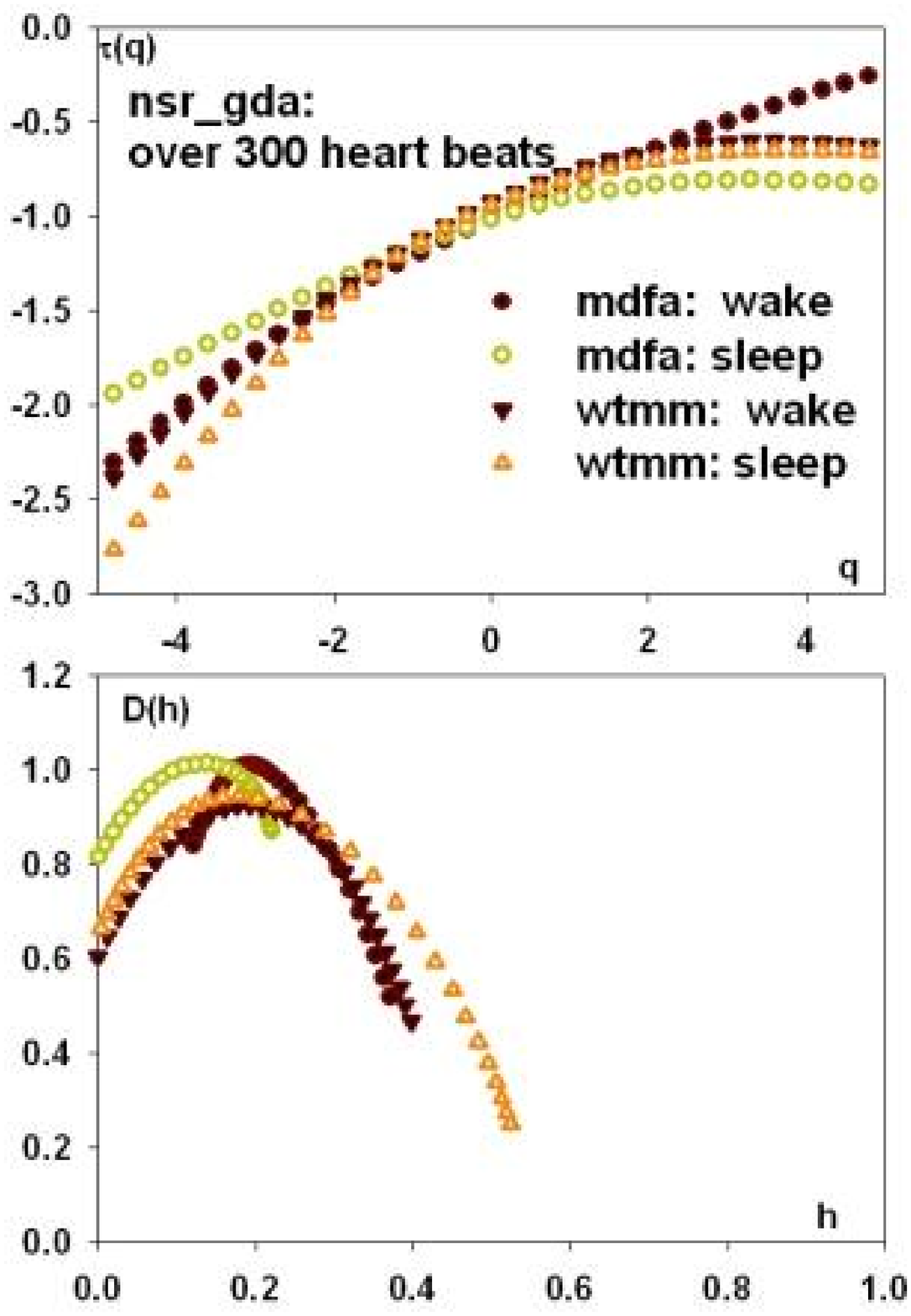}  
\includegraphics[width=0.23\textwidth]{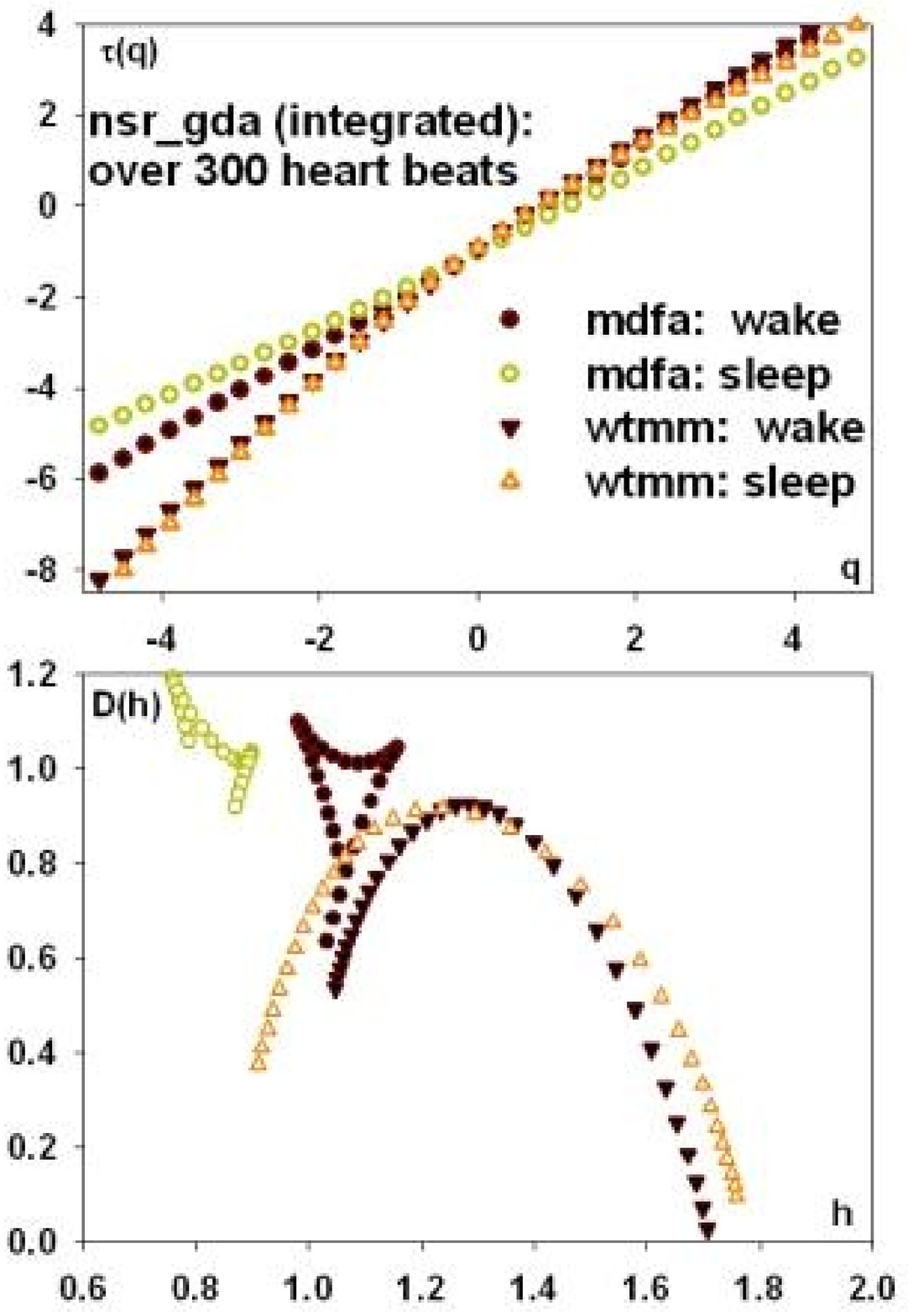}  
\caption{\small The structure functions  and spectra obtained  when the linear fits are for heart beat intervals physiologically motivated. The intervals are described in plot titles. Every third points are plotted to emphasize centers of spectral point concentration.} 
\label{spectra}
\end{figure}

In the following  we search for the best linear fits of partition functions in the particular time intervals. These intervals, transfered into number of beats, are directly related to the physiology of the heart rate control and the classical spectral analysis as it was described in Sec II. The group structure functions and the group spectrum plots are presented as follows, see Fig.\ref{spectra}:
\begin{itemize}
 \item 10--25 heart beats that cover times from 8  to 20 seconds, what allows us to estimate early compensatory mechanism  represented by LF. 
 \item  25--300 heart beats that last from 20 seconds to 4 minutes, what examines the activity of rennin-angiotensin-aldosterone system and other neurohormones, which additionally activates the sympathetic nervous system and corresponds to  VLF.
 \item  over 300 heart beats, i.e., times over 4 minutes allow to study  frequencies smaller than 0.003Hz, hence we are within ULF component.
\end{itemize}
 
We can not enter into the high frequency HF range because of numerical instabilities occurring when $q<0$.
 
It appears that in  short times, less than 20 seconds,  each method locates  the multifractal spectra in  different intervals of $h$. Both pure spectra: wake and sleep  are in $h \in (0.5, 1.0)$ in case of MDFA while maxima of WTMM spectra are about $h=0.2$ though there is a noticeable center of WTMM wake spectrum points at $h=0.6$. The spectra obtained from integrated series  sustain this observations ---  WTMM plots  are significantly separated from plots obtained by means of MDFA. It denotes that roughness of signals measured in scales smaller than  25 beats (a half of a box from Fig.\ref{methods}) is considerable differently estimated by both methods. Yet both methods indicates on presence of strong persistence in series, because the spectra contain singularity exponents with $h>0.5$. The persistence is usually related to burstiness or clustering in data --- fluctuations larger or smaller than a given level  happen subsequently, i.e.,  much often than in the case of independent observations. 

Heart rate in LF range is strongly influenced by oscillations in blood pressure. Basically the baroreflex is the modulator of RR interval in a healthy subject. However there are receptors which control the other aspects of proper blood supply, and  which activity can be evaluated by the level of randomness in LF range \cite{Sandercock}.    
Both multifractal methods produce wide spectra what can be attributed to a large variety of controlling elements under  which a cardiovascular system maintains blood supply during daily human activity. Moreover, the randomness point $h=0.5$ belongs to both spectra and it could provide an indicator of the level of the baroreflex sensitivity. MDFA plots of integrated series are moved toward a multifractal range which  corresponds to a series of integrated random walk. Thus  RR intervals are seen as rather independent of each other. However WTMM provides plots with maxima highly occupied  at about $h=0.85$ for  sleep series and around $h=0.95$ in case of wake series.

DFA picture of VLF range, Fig.\ref{spectra} plots of middle panel, again provides a huge variety of scaling exponents in case of daily activity $h\in (0,0.6)$ with two centers where spectrum points are densely located: at maximum $h=0.3$ and at $h=0.6$. WTMM spectrum in VLF range gives the wake spectrum narrower and mainly concentrated at its left wing, i.e., for  $h<0.2$. Thus comparing the two methods one can say that WTMM enhances influence of large oscillations  which are contained in $\tau(q)$ when $q>0$, while MDFA amplifies role of small fluctuations expressed by $\tau(q)$ with $q<0$. Both wake spectra  are  wider than the spectra of sleep series and the sleep spectra are located  inside the  corresponding wake spectra. Thus one can conjecture loosing variability in heart rate (measured by  a pool of scaling exponents) during the night rest in VLF range. 

The integrated wake and sleep spectra corresponding to VLF component exhibit noticeable separation  from each other: the sleep spectra are moved to left, to lower H\"older exponents, namely,  $h\in(0.6,1)$. Both sleep curves are characterized by densely occupied left wings. Hence the roughness of night series is significantly wilder. The main parts of the wake spectra occupy the similar range of $h$, i.e., $h\in(1,1.2)$. This means that WTMM wake spectrum is moved right by 1 when compared it to the pure series result but MDFA one not. However there is a spectrum center at $h=1.5$.

The third frequency component UHF corresponds to the time range where long range dependences are  usually searched. Therefore it is not a surprise that our results recalls results obtained by others, see \cite{Ivanov,filter,Kotani,Makowiec}. In case of pure series all spectra are characterized by similar set of singularity exponents $h$ concentrated at $h=0.2$ and the widths of WTMM spectra are wider than the corresponding MDFA ones. Moreover, MDFA results received  for integrated series reveal global properties known thanks to DFA studies \cite{DFA}: the healthy heart long range correlations are characterized by $\alpha_{DFA}=1$ for a daily activity  and by $\alpha_{DFA}=0.8$ during a nigh rest. WTMM spectra for integrated series differentiates day and night series by spectral width: the sleep spectrum is significantly wider than the wake one. However, the left wings of these spectra - consisting of many points, are located at the same $h$ as it is found by DFA.

\section{Conclusions}
Many naturally occurring empirical time series violate the assumption of independence of  subsequent elements. Instead the data often point at the presence of strong temporal correlations. Relating these correlations with dynamic properties of a system under study is challenging. 

Heart rate variability is a measure  of fundamental aspects of a human physiology. It is known that frequency domain analysis of heart rate provides  predictors of mortality in chronic heart failure --- LF is a predictor for  sudden cardiac death while VLF and ULV are predictors for all-causes mortality, see \cite{Guide,Frenneaux,Jong,Sandercock}.

In the following, basing on the descriptive rather aspect of multiscaling, we have analysed scaling properties of  average multifractal partition functions in these physiologically grounded interbeat intervals: LF, VLF and ULF, following normal RR intervals of  39 healthy subjects. The data was grouped according to the circadian rhythm (wake and sleep series), the methods of multifractal analysis (WTMM and MDFA approach), and  RR intervals were compared to integrated RR intervals. By applying different numerical methods we wanted to determine some objective properties hidden in series considered which discriminate  daily heart activity from its nocturnal rest. In general, we observed loss of  heart rate variability during night in VLF range and its increase in time corresponding to ULF component.

However, we have also found that details of multifractal properties such as spectrum width and location are different for different numerical methods.  A partition function obtained by WTMM  remembers about the strength of oscillations in a signal read in finer scales while MDFA provides at each scale an independent partition function. Therefore the relations between scales are more consistent with each other in case of WTMM spectra seen in different scales. It is specially noticeable in case of integrated series. The study of integrated  RR-intervals has put a significant impact in evaluation  of the participation of independent stochastic  sources. We hope that all these differences would lead to additional tools allowing to  quantify and qualify neural and humoral control of heart rate.

\bigskip
{\noindent\bf Acknowledgments}
This work was supported by the Rector of Gdansk University --- project: BW$\slash$5400-5-0220-7


\newpage
\begin{thebibliography}{99}

\bibitem{Guide}
Heart Rate Variability. Standards of measurement, physiological interpretation, and clinical use. {\it Eur.Heart J.} {\bf 17}, 354 (1996)

\bibitem{fractal}
M.Kobayashi and T.Musha, 1/f fluctuation of heartbeat period.{\it IEEE Trans.Biomed.Eng.} {\bf BME-29} 456--457 (1982);
C. K. Peng, J. Mietus, J. Hausdorff, S. Havlin, H. E. Stanley, and A. L. Goldberger, Long-Range Anticorrelations and Non-Gaussian Behavior of the Heartbeat {\it Phys.Rev.Lett.} {\bf 70} 1343 (1993);
Y.Yamamoto and R.L.Hughson, On the fractal nature of heart rate variability in humans: effects of data length and beta-adrenergic blockade. {\it Am.J.Physiol.} {\bf 266} R40-R49(1994)

\bibitem{FracMath}
K.J. Falconer,
{\it Fractal Geometry - Mathematical Foundations and Applications} John Wiley, Second Edition, 2003;

\bibitem{Riedi}
R.H. Riedi. Multifractal Processes. pp 625--716, in {\it Long-range Dependence: Theory and Applications} editors: P. Doukhan, G. Oppenheim and M. S. Taqqu, Cambridge Ma, Birkh\"auser, 2001;
W.Willinger, V.Paxson, R.H.Riedi, M.S.Taqqu pp1--38 in {\it Long-range Dependence: Theory and Applications} editors: P. Doukhan, G. Oppenheim and M. S. Taqqu, Cambridge Ma, Birkh\"auser, 2001.

\bibitem{DayNight}
L.A.N. Amaral, P.Ch. Ivanov, N. Aoyagi, I. Hidaka, S. Tomono,  A. L. Goldberger,
H.E. Stanley and Y. Yamamoto, Behavioral- independent features of complex heartbeat dynamics. {\it Phys. Rev. Lett.} {\bf 86} 6026--6029 (2001);
L.A.N. Amaral, D.J.B. Soares, L.R.da Silva, L.S.Lucena, M.Saito, H.Kumano, N.Aoyagi and
 Y. Yamamoto, Power law temporal auto-correlations in day-long record in human physical activity and their alternation with disease. {\it Eurphysics Lett.} {\bf 66}(3) 448-454 (2004)

\bibitem{Kiyono2005}
K.Kiyono, Z.R.Struzik, N.Aoyagi, F.Toga and Y. Yammamoto. Phase Transition in a Healthy Human Heart Rate.
{\it Phys.Rev.Lett.} {\bf 95} 058101 (2005)

\bibitem{BacryMuzyArnedo}
E. Bacry, J. F. Muzy and A. Arnedo.  Singularity spectrum of fractals signals from wavelet analysis: Excact results. {\it J. Stat. Phys.} {\bf 70} 635-374 (1993)\\
J.F. Muzy, E. Bacry and A. Arneodo. Multifractal formalism for fractal signals: The structure-function approach versus the wavelet-transform modulus-maxima method.
{\it Phys. Rev. E} {\bf 47} 875 (1993).

\bibitem{MDFA}
J.W. Kantelhardt, S.A. Zschiegner, E. Koscielny-Bunde, S. Havlin, A. Bunde and H.E. Stanley. Multifractal detrended fluctuation analysis of nonstationary time series.
{\it Physica A} {\bf 316} 87 (2002).


\bibitem{Lashermes}
B.Lashermes, P.Abry and P. Chainais. New insighths into the estimation of scaling exponents.
{\it Int.J.of Wavelets, Multiresolution and Information Processing} {\bf 2} No 4 497--523 (2004)
\bibitem{Oswiecimka}
P.O\'swi\c ecimka, J.Kwapie\'n,  and S. Dro/.zd/.z, Wavelet versus detrended fluctuation analysis of multifractal structures. {\it Phys.Rev.E} {\bf 74} 016103-1 (2006)

\bibitem{Jackson}
G.Jackson, C.R.Gibbs, M.K.Davies, and G.Y.H.Lip. ABC of heart failure.  Pathophysiology.
 {\it British Med.J.} {\bf  320} 167--170 (2000)

\bibitem{Malik}
M.~Malik, Heart Rate Variability and Baroreflex Sensitivity in Cardiac Electrophysiology in D.~P.~Zipes, J.~Jalife (Eds.) From Cell To Bedside, 4th edition, Saunders, 2004

\bibitem{Gregory}
M.E. Gregory, R.J. Cody, E.Nuziata and P.F. Binkley. Early Left Dysfunction Elicits Activation of Sympathetic Drive and Attenuation of Parasympathetic Tone  in in Paced Canine Model of Congestive Heart failure. {\it Circulation} {\bf 92} 551 (1995)

\bibitem{Frenneaux}
M.P.Freneaux. Autonomic changes in patients with heart failure and in post-myocardial inf raction patients. {\it Heart} {\bf 90} 1248-1255 (2004)

\bibitem{Jong}
M.J.De Jong and D.C.Randall. Heart rate variability analysis in the assessment of autonomic function in heart failure. {\it J.Cardiovasc. Nurs.} {\bf 20} 186--196 (2005). 

\bibitem{Sandercock}
G.R.H. Sandercock and D.A. Brodie, The Role of heart Variability in Prognosis for Different Modes of Death in Chronic Heart Failure.
{\it Pacing Clin. Electrophysiology} {\bf 29(8)}, 892-904  (2006)


\bibitem{Ivanov}
P.Ch. Ivanov, M.G. Rosenblum, C.-K. Peng, J. Mietus, , S. Havlin,  H.E. Stanley and  A.L. Goldberger, Scaling behaviour of heartbeat intervals obtained by wavelet-based time-series analysis. {\it Nature} {\bf 383} 323 (1999);\\
P.Ch. Ivanov, L.A.N. Amaral, A.L. Goldberger, S. Havlin, M.G. Rosenblum, H.E. Stanley and Z.R. Struzik. From 1/f noise to multifractal cascades in heartbeat dynamics.
{\it Chaos} {\bf 11} 641 (2001)

\bibitem{filter}
P.Ch. Ivanov, L.A.N. Amaral, A.L. Goldberger, S. Havlin, M.G. Rosenblum, Z.R. Struzik and H.E. Stanley. Multifractality in human  heartbeat dynamics.
{\it Nature} {\bf 399} 461 (1999)

\bibitem{Kotani}
K.Kotani, Z.R.Struzik, K. Takamasu, H.E.Stanley  and Y. Yamamoto. Model for complex heart rate dynamics in health and diseases {\it Phys.Rev.E} {\bf 72} 041904 (2005)

\bibitem{Makowiec}
D.Makowiec, R.Ga\l \c aska, A.Dudkowska, A.Rynkiewicz and M.Zwierz. Lon-range dependencies in heart rate signals - revisited. {\it Physica A} {\bf 369} 632--644 (2006)


\bibitem{Physica04}
Y. Ashkenazy, S. Havlin, P. Ch.Ivanov, C.-K Peng, V. Schultte-Frohlinde and H. E. Stanley. Magnitude and sign scaling in power-law correlated time series. {\it Physica A} {\bf 323} 19--41 (2003);
P. Ch. Ivanov, Z. Chen, K.Hu and H. E. Stanley. Multiscale aspects of cardiac control.
{\it Physica A} {\bf 344} 685 (2004)

\bibitem{Struzik}
Z.R. Struzik J. Hayano, S. Sakata, S. Kwak, and Y. Yamamoto.  1/f Scaling in Heart rate Requires Antagonistic Autonomic Control. {\it Phys.Rev.E} {\bf 70} 050901(R) (2004)

\bibitem{MeyerStiedl}
M. Meyer and O. Stiedl, Self-affine fractal variability of human heartbeat interval dynamics in health and disease. {\it Eur. J. App. Physiol.}{\bf 90} 305--316 (2003)


\bibitem{ArnedoBacryMuzy}
A. Arnedo, E. Bacry  and J. F. Muzy. The Thermodynamics of fractals revisited with wavelets.
{\it Physica A} {\bf 213} 232 (1995)

\bibitem{DFA}
C.-K. Peng, S.V. Buldyrev, S. Havlin, M.Simons, H.E. Stanley and A.L. Goldberger. Quantification of scaling exponents and crossover phenomena in nonstationary heartbeat time series. {\it Phys.Rev.E } {\bf 49 } 1685 (1995)


\bibitem{Roach}
D.Roach, A.Sheldon, w.Wilson et al.Temporarally localized contributions to measures of large-scale heart rate vatriability {\it Am.J.Physiol.} {\bf 274} H1465-71 (1998)

\bibitem{physionet}
A.L. Goldberger, L.A.N.  Amaral, L. Glass, J.M. Hausdorff, P.Ch. Ivanov, R.G. Mark, J.E. Mietus, G.B. Moody, C.-K. Peng, H.E. Stanley. PhysioBank, PhysioToolkit, and PhysioNet: Components of a New Research Resource for Complex Physiologic Signals. Circulation 101(23):e215-e220 [Circulation Electronic Pages;  $http://circ.ahajournals.org/cgi/content/full/101/23/e215 $ ]; 2000 (June 13).


\bibitem{amgdata}
Series considered can be downloaded  from {\it http://iftia9.univ.gda.pl/\~danka/DATA}


 
\end{thebibliography}
\end{document}